# Strain-induced semiconductor to metal transition in $MA_2Z_4$ bilayers

Hongxia Zhong[1], Wenqi Xiong[1], Pengfei Lv[1], Jin Yu[2,3*], and Shengjun Yuan[1*]

[1]School of Physics and Technology, Wuhan University, Wuhan, 430072, People's Republic of China

[2]Institute for Molecules and Materials, Radboud University, Heijendaalseweg 135, NL-6525 AJ Nijmegen, The Netherlands

[3]School of Mechanics and Engineering Science, Shanghai University, Shanghai, 200444, China

**Abstract**

Very recently, a new type of two-dimensional layered material $MoSi_2N_4$ has been fabricated, which is semiconducting with weak interlayer interaction, high strength, and excellent stability. We systematically investigate theoretically the effect of vertical strain on the electronic structure of $MA_2Z_4$ (M=Ti/Cr/Mo, A=Si, Z=N/P) bilayers. Taking bilayer $MoSi_2N_4$ as an example, our first principle calculations show that its indirect band gap decreases monotonically as the vertical compressive strain increases. Under a critical strain around 22%, it undergoes a transition from semiconductor to metal. We attribute this to the opposite energy shift of states in different layers, which originates from the built-in electric field induced by the asymmetric charge transfer between two inner sublayers near the interface. Similar semiconductor to metal transitions are observed in other strained $MA_2Z_4$ bilayers, and the estimated critical pressures to realize such transitions are within the same order as semiconducting transition metal dichalcogenides. The semiconductor to metal transitions observed in the family of $MA_2Z_4$ bilayers present interesting possibilities for strain-induced engineering of their electronic properties.

**Introduction**

The family of two-dimensional (2D) materials is attracting great interest since the exfoliation of monolayer graphene from graphite.[1, 2] They have many unique properties and distinguished performance for applications including mechanic engineering, electronics, information and energy technologies.[3-8] Motivated by this, many new 2D materials have been continuously proposed by theorists with excellent physical or chemical properties.[9-13] However, only few of them are synthesized successfully in monolayers,[14-19] and they are mostly lamellar in their natural bulk forms, with the component layers assembled by weak van der Waals interaction. Recently, the research enthusiasm has been extended to those 2D materials whose bulk forms are not layered in nature. A typical example is the hexagonal MXenes family, which has been fabricated by etching out the A layers from MAX materials.[20-22] In a very recent work, $MoSi_2N_4$, as a member of another new family of 2D molybdenum nitride,[23, 24] was successfully grown by chemical vapor deposition method.[20] It is reported to be semiconducting with high strength and remarkable stability.[20] Since multilayer $MoSi_2N_4$ are stacked by very weak van der Waals interaction, they are structurally flexible and easy to deform. Meanwhile, it is known that this kind of deformation may modify the physical properties of layered 2D materials dramatically. For example, in AB-stacked bilayer graphene, applying a perpendicular electric field can induce a gap opening in the electronic structure, and a subsequent compression of the interlayer spacing can widen the field-induced band gap (a 10% compression enhances the gap by 80%).[25] Another example is layered $MoS_2$, which has been shown theoretically and experimentally that it undergoes a transition from semiconductor to metal with vertical pressure.[26, 27] In general, decreasing the interlayer spacing may induce charge redistribution in layered 2D materials, it is therefore interesting and important to figure out whether the semiconducting layered $MoSi_2N_4$ and other members in the $MA_2Z_4$ (M=Ti/Cr/Mo, A=Si, Z=N/P) family are tunable under vertical strain.

In this work, we will study this problem theoretically from first-principle density-functional theory calculations. Indeed, we show that the compressive strain can serve as an effective tool to tailor the electronic properties of bilayer $MoSi_2N_4$. Importantly, a semiconductor to metal transition is observed when the vertical compressive strain reaches 22%. Such a semiconductor to metal transition is a result of opposite energy shifts of states in two layers. The opposite energy shifts are further attributed to the enhanced interlayer interaction via asymmetric charge redistribution at the interface. We further investigate the threshold vertical compressive strain and estimate the corresponding pressure for $MoSi_2N_4$ and other semiconducting $MA_2Z_4$ bilayers. Our theoretical studies of their tunable electronic properties provide a guide for further experimental investigation, and bring opportunities to build novel electro-mechanical devices from these newly synthesized 2D materials.

**Calculation method**

Our calculations are performed using the projector augmented wave (PAW) method[28] implemented in the Vienna ab initio simulation package (VASP) code.[29] Perdew, Burke, and Ernzerhof (PBE) form of the generalized gradient approximation (GGA) exchange correlation functional[30] with van der Waals corrections (vdW-DFT),[31, 32] and the PAW pseudo potentials[28] are adopted. The cut-off energy is set to 500 eV after convergence tests. A Γ-centered Monkhorst-Pack $k$-point grid[33, 34] of 9×9×1 is used for relaxations and the grid of 15×15×1 for property calculations. In our current calculations, the total energy is converged to less than $10^{-6}$ eV, and the maximum force is less than 0.01 eV/Å during the optimization. A vacuum space along $z$-axis is larger than 20 Å to avoid spurious interactions. During the structure relaxation, the $z$ coordinates of N atoms at the interface are fixed to ensure the layer spacing, and the $z$-axis is fixed to ensure the thickness of the vacuum slab. The spin–orbit coupling (SOC) have been examined for unstrained bilayer sheets.

**Results and Discussions**

We first perform calculations on monolayer $MoSi_2N_4$ and compare with reported results. The lattice constant is 2.91 Å, and the indirect band gap is 1.789 eV. They are in agreement with previous values of 2.94 Å and 1.744 eV, respectively.[20] With different stacking configurations, two $MoSi_2N_4$ monolayers can form three different bilayer structures (AA, AB and AC) with high symmetry.[20] Based on our calculated total energies, the order of stability of these three structures is AA < AB < AC. Hence, we take most stable AC stacking to discuss the related electronic properties in the rest of the paper.

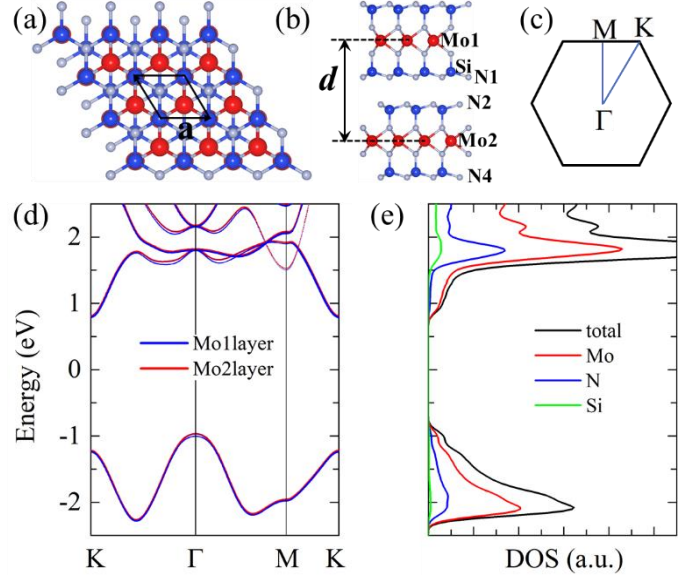

Fig. 1: Top (a) and side (b) views of atomic structure of bilayer $MoSi_2N_4$. The primitive cell and lattice vector **a** are indicated in (a). $d$ is the interlayer distance between the top Mo1 and bottom Mo2 sublayers. (c) The first Brillouin zone of bilayer $MoSi_2N_4$. Projected band structure (d) and partial density of states (e) of bilayer $MoSi_2N_4$. The band gap center is set to be zero.

The ball-stick structures of AC-stacked $MoSi_2N_4$ are presented in Fig. 1. The unit cell and unit vectors are marked by black vectors. As shown in Figs. 1(a) and 1(b), the Si atoms of the top monolayer are superimposed on the Mo2 atom of the bottom monolayer, and the Mo1 atom of the top monolayer are superimposed on the N1 and N4 atoms of the bottom monolayer, i.e., sitting above the hexagon centers of the bottom lattice. The optimized lattice constant is 2.91 Å, which is the same as that of monolayer. The layer distance between Mo1 and Mo2 atoms is 11.030 Å, much larger than the one (6.23~ 6.54 Å) of bilayer transition metal dichalcogenides (TMDCs), such as $MoS_2$, $WS_2$, $MoSe_2$ and $WSe_2$.[35, 36] This suggests that the interlayer vdW interaction in bilayer $MoSi_2N_4$ will be much weaker than bilayer TMDCs. Such interlayer interaction is too weak so that the band structure of bilayer $MoSi_2N_4$ is almost the same as two isolated monolayers (as shown in Fig. 1(d)), except that there are negligible splittings of the highest (lowest) occupied (unoccupied) states along Γ-K and Γ-M directions. For bilayer $MoSi_2N_4$, the valence band maximum (VBM) locates at Γ point, about 0.271 eV higher than the highest occupied states at K point, while the conduction band minimum (CBM) is at K point, indicating an indirect band gap of 1.761 eV. If we include the spin-orbital coupling (SOC) in the calculations, there will be a band splitting (0.141 eV) of the highest occupied states at K point. This spin-induced splitting is compared to the value (0.149 eV) of $MoS_2$.[37, 38] By contrast, no spin splitting is observed in the CBM at K point after including the SOC effects. In Fig. 1(e), the projected density of states indicates that both VBM and CBM are dominated by Mo atoms, with partly contributions from N atoms, but almost no contribution from Si atoms. By a more detailed analysis of the orbital contributions, the band edges are indeed mainly from the localized Mo $dz^2$ orbitals. It is worth to note that the observed SOC effects and orbital characters of the states at band edges are very similar to TMDC monolayers. Apart from the valence band splitting at K point, the SOC effect does not modify the band dispersion and band gap of bilayer $MoSi_2N_4$, and thus we will neglect it in the following sections.

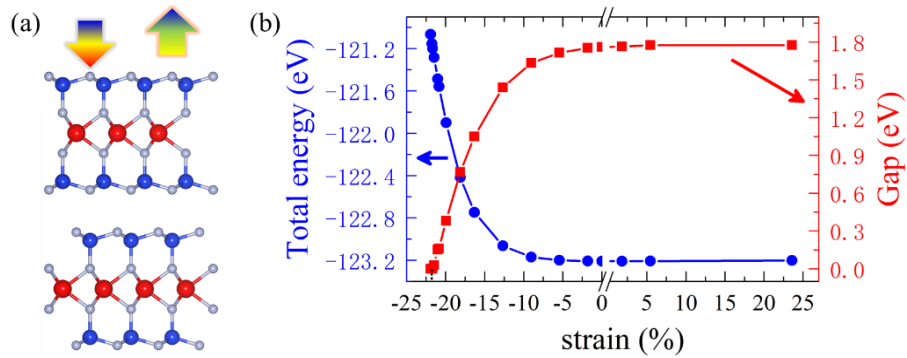

Fig 2: (a) Schematic diagram of the bilayer $MoSi_2N_4$ under an external vertical strain along the direction perpendicular to the layer, where the arrows pointing up and down represent a stretched and compressive strain, respectively. (b) The variation of total energy and band gap of bilayer $MoSi_2N_4$ as function of the interlayer spacing. The zero band gaps have been indicated by the black dotted line.

Applying vertical strain is an effective way to modulate the electronic properties of layered materials. It can be realized by hydrostatic pressure,[39] vacuum thermal annealing, nanomechanical pressures, or via inserting hexagonal BN dielectric layers.[40] In our calculations, the vertical strain is implemented by the change of the interlayer distance and characterized by a quantity $\delta = (d-d_0)/d_0$, where $d$ and $d_0$ is the vertical distance for strained and unstrained bilayers. Fig. 2(a) shows schematic diagram of bilayer $MoSi_2N_4$ under vertical strain perpendicular to the plane of the layer, where the arrows pointing upward and downward indicate a stretched and compressive strain, respectively. Fig. 2(b) depicts the evolutions of the total energy and band gap of bilayer MoSi$_2$N$_4$ under different vertical strains. For the intrinsic structure ($\delta$ = 0), bilayer MoSi$_2$N$_4$ with interlayer distance 11.030 Å is the most stable configuration with the lowest total energy. Within the range $\delta \geq$ -0.09, the total energy keeps almost the same, indicating that the interlayer interaction is too weak to affect the electronic properties. While for the vertical strain smaller than -0.09, i.e., a compressed strain more than 9%, the total energy increases monotonically with the decreasing vertical strain. Thus, the vertical strain with $\delta$=-0.09 can be regarded as the turning point where the interlayer interaction becomes so strong that cannot be neglected.

For the change of the band gap, when the value of $\delta$ is above the threshold -0.09, the indirect band gap is almost unchanged, similar to the property of total energy. When $\delta$ is below the threshold -0.09, the band gap decreases significantly. This threshold strain corresponds to a vertical distance of 3.02 Å between inner N atoms of two monolayers, which is comparable to the layer spacing of layered TMDCs,[41] black phosphorene (3.21 Å),[42] and borophene (3.07 Å).[43] In layered TMDCs, the interlayer interaction does affect their electronic structures greatly, changing the band gap from direct in monolayer to indirect in multilayers.[19, 44, 45] Here, if we continuously increase the compressed strain in bilayer MoSi$_2$N$_4$, an electronic phase transition from semiconductor to metal is finally observed when the value of $\delta$ reaches -22%. The corresponding vertical compressive pressure P is 18.66 GPa, which is estimated in terms of $P = \frac{E-E_0}{(d_0-d)*A}$,[41, 46] where $E$ and $E_0$ are total energies at the layer distance $d$ and $d_0$ (equilibrium), and A is the area of the cell. Such a transition pressure is on the similar order of bilayer TMDCs (5.10-16.28 GPa).[41] Generally, the electronic phase transition is accompanied by structural transition under high pressure.[47] While for bilayer MoSi$_2$N$_4$ under the transition pressure around 18.66 GPa, AC stacking is still found to be the most stable one, indicating that there will be no structural transition during the strain process. This is consistent with the excellent ambient stability observed in layered MoSi$_2$N$_4$.[20] Our predicted transition pressure of bilayer MoSi$_2$N$_4$ can be realized by hydrostatic pressure,[39] making it attractive for electro-mechanical applications.

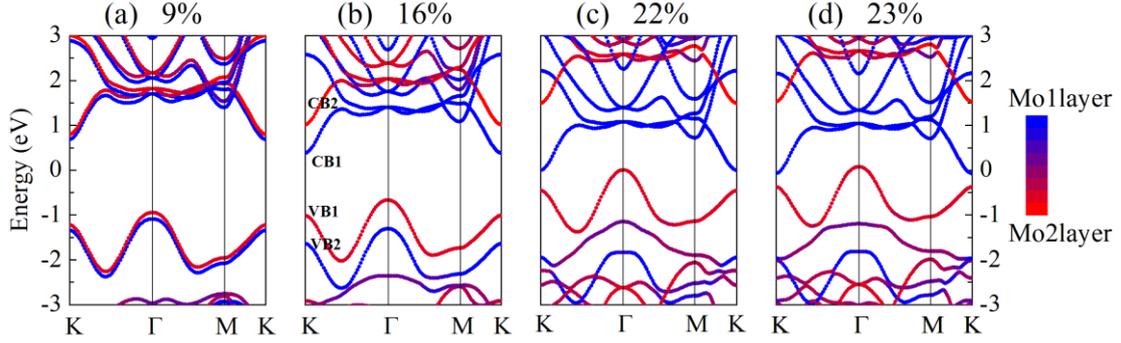

Fig 3: Band structures of bilayer $MoSi_2N_4$ under vertical compressive strains from 9% to 23%. All bands are shifted to align the deep Mo 4$s$ state. Two conduction bands (CB1 and CB2) and two valence bands (VB1 and VB2) are labeled in (b).

Because the band gaps vary significantly within the range $\delta \leq -0.09$ in Fig. 2(b), we only focus on the vertical compressive strains in the following. In Fig. 3, we present the band structures of bilayer $MoSi_2N_4$ under four representative compressive strains from 9% to 23%. As the result of $\delta = 9\%$ shown in Fig. 3(a), notable changes appear in the electronic structure of bilayer $MoSi_2N_4$ that there are slight splittings of energy bands. These band splittings become more evident with the increasing of the compressive strain (see the changes of $\delta$ from 9% to 16% and 22%). Here, we can quantitatively define the rigid energy shift as the energy difference between the VB1 (CB1) and VB2 (CB2) at K ($\Gamma$) point as labeled in Fig. 3(b). When the compressive strain increases from 9% to 23%, the energy shift increases significantly from 0.116 to 1.605 eV. More importantly, the energy dispersion of each individual band keeps almost the same during the compression process. As a result, the charge carrier will maintain its small effective mass and high mobility (270-1200 $cm^2\ V^{-1}s^{-1}$), similar as their values in pristine monolayer. If we inject electron or hole into a strained bilayer, crossing the energy band CB1 or VB1, then only the Mo1 layer or Mo2 layer will be conducting. Carrier injection thus can be used to effectively tune the transport properties of stained $MoSi_2N_4$ bilayers. It is also noted that the work function of bilayer $MoSi_2N_4$ is almost unchanged under different vertical stains. On the other hand, with the increasing compressive strain, both VBM and CBM are driven continuously toward the Fermi level, and thus reduce the energy band gap. The change of the band structure show that the indirect band gap of bilayer $MoSi_2N_4$ decreases from 1.633 to 0 eV, when the compressive strain increases from 9% to 22%. After the compression reaches the critical value 22%, the metallic feature appears in the band structure as shown in Figs. 3(c) and 3(d) where δ ≥ 22%, indicating clearly a robust semiconductor to metal phase transition.

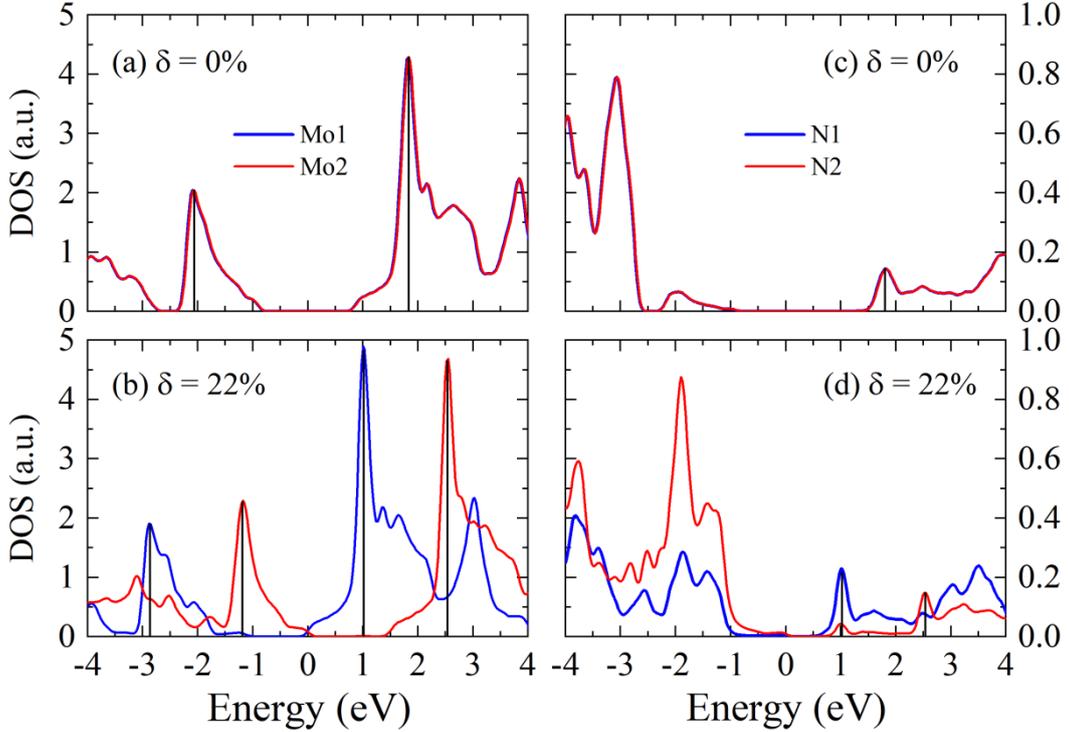

Fig 4: Partial density of states (PDOS) of Mo1, Mo2, N1 and N2 atoms (as labeled in Fig. 1) in bilayer $MoSi_2N_4$ with or without strain. All the energies are shifted to align the deep Mo 4$s$ state.

In order to find the origin of this semiconductor to metal phase transition, we extract the partial density of states (PDOS) of bilayer $MoSi_2N_4$ with or without strain. As we have already shown that the band edges of bilayer $MoSi_2N_4$ are dominated by Mo states, we thus focus on the Mo states in two layers. In Fig. 4(a), the calculated PDOS of two Mo atoms from different layers in the pristine bilayer $MoSi_2N_4$ are totally degenerate, because of the symmetric atomic structures of the two layers. Under a vertical compressive strain 22%, the overall shape of the PDOS from two Mo are preserved but they have opposite energy shift, and there is an overlap of highest valence state of one Mo and the lowest conduction state of another Mo, as shown in Fig. 4(b). This overlap of Mo states agrees with the semiconductor to metal transition in bilayer $MoSi_2N_4$ under the critical strain of 22%. If we make an analysis of the PDOS from N atoms in different layers, although their contributions to the states at the band edges are much smaller, the spectrum shift of each N atom is similar to the Mo atom within the same layer. Furthermore, there is a significant enhancement of PDOS of N1 and N2 atoms, indicating that there is an enhancement of interlayer interactions between the $p_z$ orbitals of N atoms crossing the interface of the two layers.

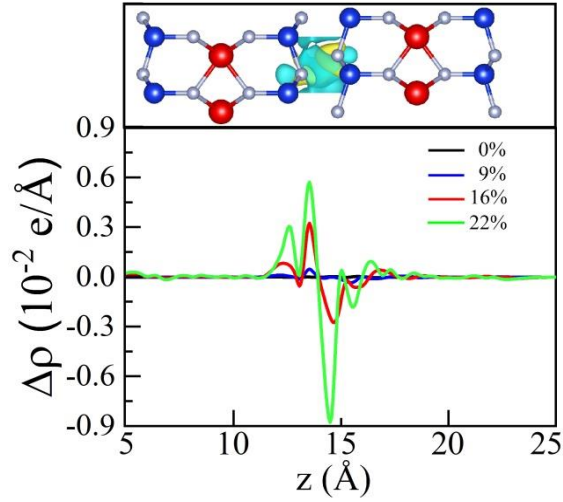

Fig 5: Plane-averaged charge density difference $\Delta\rho(z)$ of bilayer $MoSi_2N_4$ under vertical compressive strain from 0% to 22%. The top panel is a three-dimensional charge density difference under the strain 22%, and the isosurface value is 0.002 e/Å$^3$. The green and yellow areas represent electron accumulation and depletion, respectively.

The opposite energy shift of the PDOS indicates a possible strain induced change of the electric potentials on these atoms. We thus calculate the plane-averaged charge density difference $\Delta\rho(z)$ along the vertical direction (z-axis) for $MoSi_2N_4$ bilayers with different strain and plot the results in Fig. 5. Here, $\Delta\rho(z)$ is calculated by the charge density difference between bilayer and two non-interacting monolayers. A positive $\Delta\rho(z)$ indicates the accumulating of the charge density, while a negative value means the depletion. For unstrained bilayer $MoSi_2N_4$, $\Delta\rho(z)$ is almost zero for all z coordinates, suggesting no charge transfer between layers. This is also consistent with the observed very weak interlayer vdW interaction in pristine bilayer $MoSi_2N_4$. For strained bilayer $MoSi_2N_4$, we see clearly changes of $\Delta\rho(z)$ are totally different on the two sides of the interface. Although on both layers, there are fluctuations of $\Delta\rho(z)$, but clearly on one side the overall values of $\Delta\rho(z)$ are negative, and on the other side they are positive, indicating the charge transfer between the two layers. The amount of charge transfer $Q$, obtained by the integral of $\Delta\rho(z)$ along one side of the interface, increase monotonically with the strain. Furthermore, if we calculate the ratio of $Q/\Delta$gap for different strain, we obtain a similar value of 0.035. This correlation between Q and reduced energy gap indicate that, the asymmetric charge redistribution at the interface causes the opposite energy shift of the two layers, and subsequently the semiconductor to metal phase transition at the critical compressive strain.

**Table 1.** Distance between two Z atoms from different layersin pristine bilayer $MA_4Z_2$ or with a critical strain, and estimated pressure required for realizing the semiconductor to metal transition. We also list the vertical separation between interlayer anion atoms and the corresponding pressure for semiconductor to metal transition in some AB-stacked TMDCs for comparison.[41]

|          | Bilayer separation at transition (Å) | Bilayer separation at equilibrium (Å) | Transition pressure (GPa) |
|----------|--------------------------------------|---------------------------------------|---------------------------|
| $MoSi_2N_4$ | 1.62 | 4.02 | 18.66 |
| $MoSi_2P_4$ | 2.42 | 4.02 | 2.38 |
| $CrSi_2N_4$ | 2.62 | 4.03 | 2.18 |
| $TiSi_2N_4$ | 1.20 | 4.04 | 32.04 |
| $MoS_2$     | 2.14 | 3.11 | 8.52 |
| $MoSe_2$    | 2.23 | 3.19 | 8.37 |
| $MoTe_2$    | 2.69 | 3.37 | 5.10 |
| $WS_2$      | 1.80 | 3.39 | 16.28 |
| $WSe_2$     | 2.21 | 3.35 | 15.83 |

To complete our study, we studied numerically the strain-induced semiconductor to metal transition in other $MA_2Z_4$ bilayers, and summarize their structural parameters and transition pressures in Table 1. Owing to the weak interlayer interaction, the interlayer separation for the unstrained bilayer $MA_2Z_4$ is all around 4.02 Å, larger than those (3.11-3.39 Å) of bilayer TMDCs.[41] By increasing the vertical compressive strain, the interlayer separation decreases accordingly. Under the transition strain, all considered $MA_2Z_4$ bilayers change from semiconductor to metal t, and the interlayer separation ranges from 1.62 to 2.62 Å, similar to the range (1.80-2.69 Å) of bilayer TMDCs at the transition point.[41] The estimated transition pressure ranges from 2.18 GPa in bilayer $CrSi_2N_4$ to 18.66 GPa in bilayer $MoSi_2N_4$, which is on the same order of that (5.10-19 GPa) of well-studied layered TMDCs.[41, 47]. In particularly, bilayer $CrSi_2N_4$ and $MoSi_2P_4$, have very smaller transition pressures (2.18 - 2.38 GPa) comparing to other studied $MA_2Z_4$. Our calculations suggest that the strain-induced semiconductor to metal transition in semiconducting bilayer $MA_2Z_4$ can be easily realized experimentally. Furthermore, our extended calculations show that semiconductor to metal transition also appears in four-layer $MoSi_2N_4$, and the transition strain is around 18%, corresponding to 15.03GPa, which is smaller than that of stacked bilayer. Therefore,.

we expect that similar transition can be observed in all multilayer and bulk $MA_2Z_4$ with experimentally reachable pressure.

**Conclusion**

In conclusion, we have studied the electronic properties of $MA_2Z_4$ (M=Ti/Cr/Mo, A=Si, Z=N/P) bilayers under different vertical strains using first-principle density-functional theory calculations. Taking the recently fabricated $MoSi_2N_4$ as an example, our results reveal that the electronic properties of bilayer $MoSi_2N_4$ can be effectively tuned by the vertical compressive strain. To be more exact, with an increasing compressive strain, the band gap of bilayer $MoSi_2N_4$ monotonically decreases and finally close when the vertical strain reaches 22%. This is a result of the opposite energy shift of the states in different layers, which is driven by the asymmetric charge redistribution on the inner A-A sublayer at the interface. We further conformed that similar semiconductor to metal transitions exists in other strained $MA_2Z_4$ bilayers, and the estimated transition pressure to realize such transition ranges from 2.18 GPa in $CrSi_2N_4$ to 32.04 GPa in $TiSi_2N_4$. Our theoretical predictions provide a guiding for further exploring the strain-tunable electronic properties of layered semiconducting $MA_2Z_4$.

**Acknowledgement**

This work is supported by the National Key R&D Program of China (Grant No.2018YFA0305800) and National Natural Science Foundation of China (Grant No. 11947218). Hongxia Zhong acknowledges the support by the China Postdoctoral Science Foundation (Grant No.2018M640723). Numerical calculations presented in this paper have been performed on a supercomputing system in the Supercomputing Center of Wuhan University.